\documentclass[prl,twocolumn,superscriptaddress]{revtex4}
\usepackage{graphicx}
\usepackage{amsmath}
\usepackage{color}
\usepackage{array}
\usepackage{hyperref}
\usepackage{grffile}
\usepackage{float}

\begin{document}

\title{Disorder-driven metal-insulator transitions in  deformable lattices}

\author{Domenico Di Sante}
\affiliation{Institute of Physics and Astrophysics, University of W\"urzburg, W\"urzburg, Germany}
\affiliation{Consiglio Nazionale delle Ricerche (CNR-SPIN), Via Vetoio, L'Aquila, Italy}
\author{Simone Fratini} \affiliation{Institut N\'{e}el-CNRS and Universit\'{e}
 Grenoble Alpes, Bo\^{i}te Postale 166, F-38042 Grenoble Cedex 9, France}
\author{Vladimir Dobrosavljevi\'c}
\affiliation{Department of Physics and National High Magnetic Field Laboratory, Florida State University, Tallahassee, Florida 32306, USA}
\author{Sergio Ciuchi}
\affiliation{Department of Physical and Chemical Sciences, University of L'Aquila, Via Vetoio, L'Aquila, Italy I-67100}
\affiliation{Consiglio Nazionale delle Ricerche (CNR-ISC) Via dei Taurini, Rome, Italy I-00185}

\begin{abstract}
We show that in presence of a deformable lattice potential, the nature of the
disorder-driven metal-insulator transition (MIT) is fundamentally changed with
respect to the non-interacting (Anderson) scenario. For strong disorder, even a
modest electron-phonon interaction is found to dramatically renormalize the
random potential, opening  a mobility gap at the Fermi energy. This process,
which reflects disorder-enhanced polaron formation, is here given a microscopic
basis by treating the lattice deformations and  Anderson localization effects
on the same footing. We identify an intermediate "bad insulator" transport
regime which displays resistivity values exceeding the Mott-Ioffe-Regel limit
and with a negative temperature coefficient, as often observed in strongly disordered metals. 
Our calculations reveal that this behavior originates 
from significant temperature-induced rearrangements of electronic
states due to enhanced interaction effects close to the disorder-driven MIT.
\end{abstract}

\maketitle

\paragraph{Introduction.---}

Sufficiently strong disorder typically leads to the formation of bound
electronic states. This physical process -- Anderson localization -- is by now
well understood in the noninteracting limit \cite{Anderson,MirlinEvers}. Still, even
early experimental  and theoretical studies stressed \cite{LeeRMP} that omnipresent
interaction effects cannot be disregarded, although they proved difficult to
tackle. From the theoretical point of view, the pitfall of conventional
weak-coupling theories has been the challenge in incorporating the strong
interaction effects at the same level as disorder, especially in compounds with
local magnetic moments and various Mott systems. The theoretical landscape
changed dramatically following the rise of Dynamical Mean-Field Theory (DMFT)
ideas \cite{dmft96}, which provided a new perspective. Several intriguing phenomena, such as
disorder-driven non-Fermi liquid behavior \cite{RoP2005review}, glassy dynamics of electrons \cite{pastor-prl99}, and even the physics of Mott-Anderson transitions \cite{motand} have been captured, with focus on systems with strong electronic correlations. 

Many other materials, including the famous A15 compounds \cite{fisk76prl}, as well as
"phase-changing" amorphous alloys \cite{siwgriest11natphys},  can be experimentally tuned through
disorder-driven MITs \cite{mott-book90,dobrosavljevic2012conductor}, but they often do not display \cite{LeeRMP} strong electronic correlations of the Mott type \cite{motand}. In many such systems, transport on the metallic side is dominated by conventional electron-phonon scattering, leading to
familiar linear resistivity at ambient temperatures. This behavior is modified
as disorder increases, leading to a change of sign in the temperature coefficient
of resistivity (TCR), and eventually a crossover to the insulating behavior. While the
precise mechanism has long remained a puzzle  \cite{LeeRMP}, one thing is clear: The relevant
transport processes must reflect a nontrivial interplay of the dynamical lattice
deformations and disorder.

Soon after the discovery of localization, Anderson himself \cite{AndersonNat72}
suggested that in real systems lattice deformations could dramatically affect
the random potential, possibly leading to a gap opening on the
insulating side.  Ramakrishnan \cite{LeeRMP} subsequently argued that, as
soon as translational invariance is lost, a direct Hartree-type electron-phonon
interaction arises that can strongly renormalize the disorder, reminiscent of charged
impurity screening by Coulomb interactions; in contrast with the Coulomb case, however,
the lattice deformations should {\it enhance} (i.e.
anti-screen) the effects of disorder. While these early ideas
and subsequent works \cite{Shore,Cohen83,Shinozuka}
strongly emphasized the very significant role of lattice deformations in disordered materials,
so far no systematic theory has been put forward that can provide a
picture of the resulting MIT.

In this Letter we present the conceptually simplest theory of disorder-driven MITs,
treating Anderson localization at the same level as the electron-phonon
interaction. This is achieved by blending Typical Medium Theory (TMT) for
Anderson localization \cite{DobroEPL2003}, and Dynamical Mean-Field Theory \cite{dmft96} 
to tackle lattice deformations. The accuracy of the former has  been validated by
appropriate cluster extensions, showing it to capture most trends for Anderson
transitions \cite{Ekuma14,ZhangJarrell}. Careful systematic studies have also shown the DMFT
approach to the electron-phonon problem totally capable of capturing non-perturbative polaronic effects,
describing both incoherent self-trapping and coherent quasiparticle properties
\cite{depolarone,rhopolaron03,MillisA15,MillisPRB1996,CiuchiPRB1998}.
In clean systems polaron formation occurs only at very
strong coupling, uncharacteristic of typical metals. We find that the situation
is dramatically different in presence of sufficient disorder. Here, very
pronounced disorder-induced lattice deformations arise in the vicinity of the
MIT even in the most common cases
of weak/moderate electron-phonon coupling,
dominating most observables. It is precisely on such experimentally relevant region that we  
concentrate below.

\paragraph{Model and methods.---}

We study the following disordered Holstein model
\begin{equation}
H= - t\sum_{\langle i j\rangle} c^\dagger_i c_j + \sum_i \epsilon_i c^\dagger_i c_i -g\sum_i c^\dagger_i c_i X_i +H_{ph}
\label{eq:H}
\end{equation}
where $c^\dagger_i$ ($c_i$) are creation (annihilation) operators for electrons moving
on a lattice of sites $i$ with transfer integral $t$. The
site energies $\epsilon_i$ are randomly chosen from a uniform distribution
of width $2W$,
$P_0(\epsilon_i)=\theta(W^2-\epsilon_i^2)/(2W)$.
In addition to the random potential,
the electrons  interact locally  with
dispersionless phonons of frequency $\omega_0=\sqrt{K/M}$ described by $H_{ph}= \sum_i \frac{KX_i^2}{2}+ \frac{P_i^2}{2M}$.
The strength of the electron-phonon coupling is measured by the dimensionless parameter $\lambda=g^2/(2KD)$, with $D$ the
half bandwidth.
As our focus is on metals 
where electron correlations do not play a major role,
we  ignore the spin degree of freedom and consider a half-filled
band with a semi-circular density of states (DOS).

In DMFT for spatially homogeneous systems,
the lattice problem Eq. (\ref{eq:H}) is mapped onto a single impurity which is
coupled to the rest of the system via a dynamical Weiss field $G_0^{-1}$ \cite{dmft96}. The latter is determined
self-consistently
by spatially averaging the local Green's function $G$ over all the equivalent sites of the lattice.
While this theory (which in the non-interacting limit is known as the coherent
potential approximation, or CPA) can  describe certain
properties of disordered electron systems
on the average \cite{CPAvsExact,AlvermanPSSC2004}  it does not
account for the large and non-normal fluctuations that cause Anderson localization of the electronic carriers.
To this aim an alternative mean-field description can be introduced that
focuses on the most probable, or \textit{typical} quantities:
the typical density of states (TDOS) is defined as the geometric average of  the local DOS
over  sites with random energy $\epsilon$ as $\rho_{typ}(\omega)=\exp\left[ \int d\epsilon P_0(\epsilon)
\ln \rho(\omega,\epsilon) \right]$.
According to the Fermi golden rule, the escape rate from a given site can be
estimated  as  $\tau^{-1}_{esc}\simeq t^2\rho(\omega,\epsilon)$  \cite{Anderson};
the typical escape rate is therefore proportional to the TDOS, which
represents the density of mobile states at a given energy. The region in the band
where $\rho_{typ}(\omega)$ vanishes  identifies the mobility edge,
and its value $\rho_{typ}(0)$ at the Fermi energy
serves as an order parameter for the Anderson transition \cite{DobroEPL2003}.

Solving the full model Eq. (\ref{eq:H}) involves the calculation of
$\Sigma_{e-ph}(\omega,\epsilon)$, the local electron-phonon self-energy in presence of site disorder.
To this aim we first  apply the  formulation of Refs. \cite{MillisPRB1996,CiuchiPRB1998} where
the phonons are represented by a classical field that
responds self-consistently to the electrons. The advantage of this method, 
which is valid in the adiabatic limit $\omega_0/D\to 0$, is that the lattice randomness and the 
deformations are treated on the same footing, for any value of $\lambda$.
The effects of phonon quantum fluctuations for $\omega_0\neq 0$ are subsequently included
via a diagrammatic non-crossing approximation
(NCA) valid in the weak and moderate electron-phonon coupling regimes \cite{CaponeCiukPRL2003,NCACPA,Sr2TiO4}.

\begin{figure} 
\centering
\includegraphics[width=0.8\columnwidth]{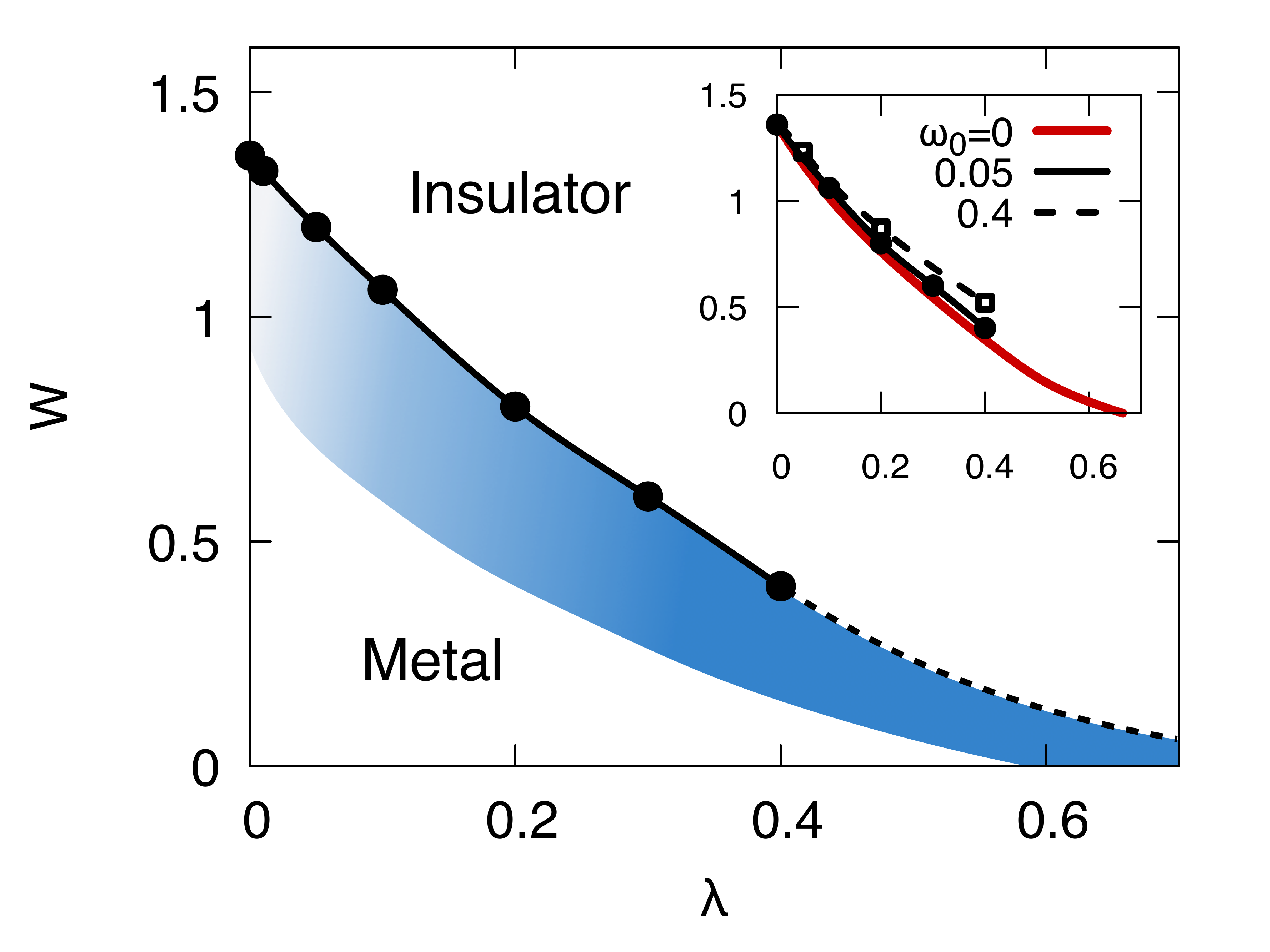}
\caption{\textit{Phase diagram.} Metal-insulator transition (MIT) at $T=0$ 
calculated for quantum phonons in the adiabatic regime ($\omega_0=0.05$).
The shaded area corresponds to the "bad insulator" behavior seen in transport (see text).
The dashed line is a sketch of the expected behavior approaching the clean limit.
The inset shows the effect of increasing phonon quantum fluctuations.}
\label{fig:PD}
\end{figure}

\paragraph{Disorder-induced polaron transition and mobility gap.---}
Fig. \ref{fig:PD}(a) shows the phase diagram obtained from the
solution of the TMT-DMFT equations.
In absence of electron-phonon interactions, $\lambda=0$, the theory 
reduces to that of Ref. \cite{DobroEPL2003}: a transition from a metal to an
Anderson insulator  occurs at a critical disorder strength $W_c^{(0)}=e/2\simeq 1.36$,
identified by  $\rho_{typ}(0)=0$ (all states are localized). Turning the electron-phonon
coupling on stabilizes the Anderson insulator, decreasing $W_c$: as anticipated,
the effect is opposite to that of repulsive Coulomb interactions
 \cite{Aguiar09} that instead screen out the effects of disorder.
As we proceed to show, polaron states characteristic of the strong coupling limit 
exist all the way down to $\lambda\to 0$,
reflecting the positive interplay of disorder and electron-phonon coupling \cite{NCACPA,SangiovanniAssaad}.

\begin{figure}
\centering
\includegraphics[width=\columnwidth]{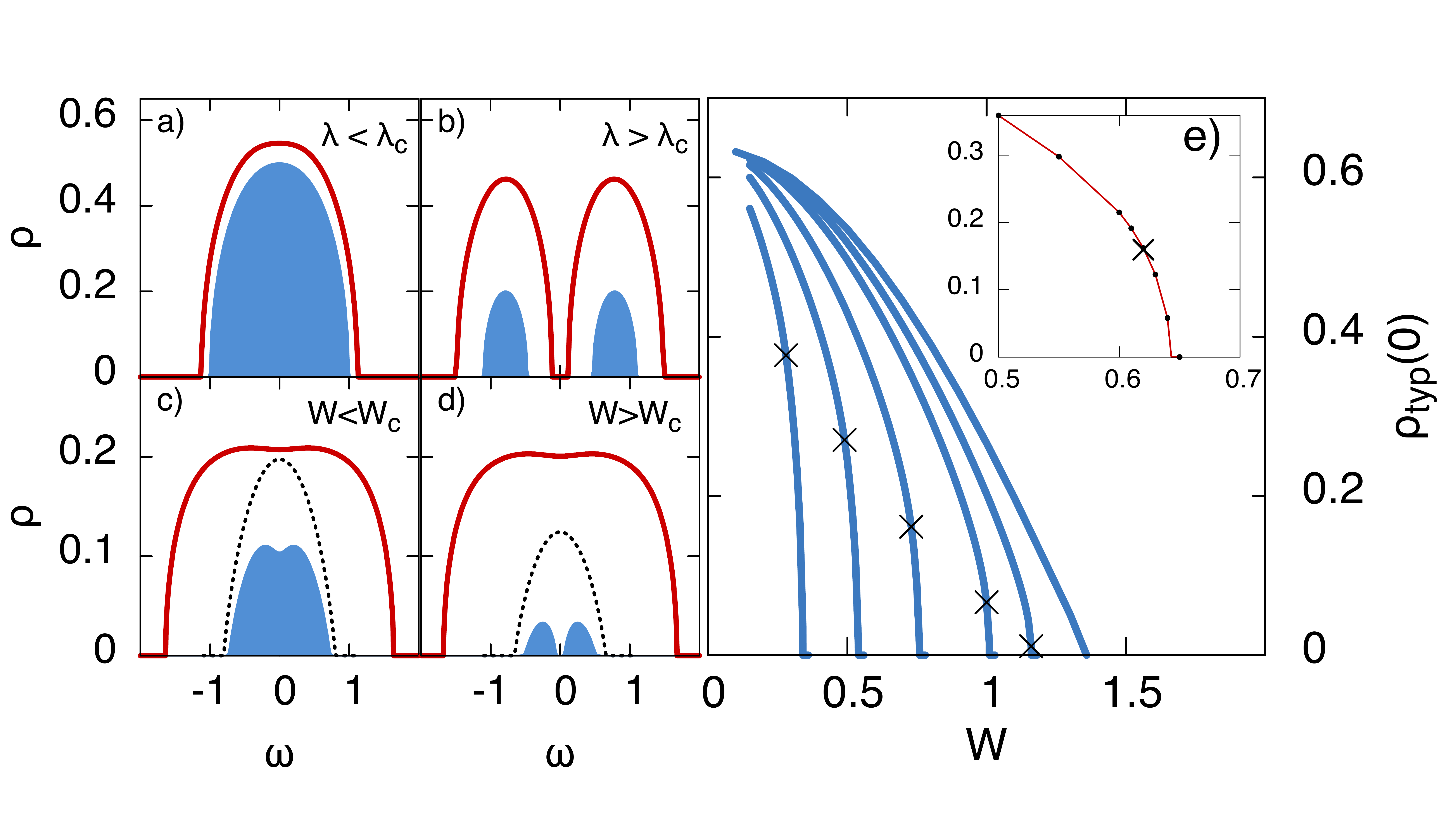}
\caption{\textit{Spectral features and order parameter.}
Panels (a-d): average (bold line) and typical DOS (shaded) across the MIT at $T=0$, 
for classical phonons.
(a,b)  at $W=0.1$, for $\lambda=0.5, 0.64$; (c,d)
at $\lambda=0.05$, for $W=1.1, 1.2$; the dotted line is the TDOS at $\lambda=0$, shown for comparison;
the DOS in the second row have been scaled down by an arbitrary factor for clarity.
(e) Order parameter vs. disorder amplitude $W$. From right to left,
$\lambda=0,0.05,0.1,0.2,0.3,0.4$.
The crosses mark the polaron transition. The inset shows the quantum case for $\lambda=0.3$ and $\omega_0=0.2$.  
}
\label{fig:DOSvsW_DOS}
\end{figure}

To illustrate the evolution of the electronic properties across the transition, we report
in Figs. \ref{fig:DOSvsW_DOS}(a-d) both the average DOS
and the TDOS, providing respectively
the spectrum of electronic states and their conductive
character. Both quantities, calculated here in the classical phonon limit (Fig. 1 inset), are accessible experimentally through 
local spectroscopic probes \cite{Yazdani}.
For strong electron-phonon interactions and weak disorder (panels a-b),
as the electron-phonon coupling strength reaches the critical value $\lambda_{c}$
a mobility gap opens at $\omega=0$ indicating the localization of states
around the Fermi energy (TDOS, shaded). This  is rapidly followed by the disappearance
of the states themselves (DOS, red), as
both phenomena are driven by polaron formation: 
self-trapping of the charges due to strong electron-phonon interactions
(and pinned by weak disorder)
leads to a binary distribution of lattice displacements that splits the excitation
spectrum into two separate subbands \cite{MillisPRB1996}.

Strikingly, the opening of a mobility \textit{gap} at the MIT persists down to the weakly interacting limit, 
a situation that is of broad relevance to many disordered materials.
The  behavior observed
as the transition line is crossed upon increasing $W$ at small $\lambda$  (cf. Fig. 2(d))
is  fundamentally different from the case where lattice effects are ignored from the beginning, where 
all states become localized at the MIT 
and no mobility gap is observed (dotted line). At variance with the strong electron-phonon coupling limit,
however, here the mobility gap  opens at the Fermi energy in an electronic spectrum that
is otherwise essentially unperturbed. 
The critical behavior of the order parameter, shown in Fig.  2(e), is also
modified accordingly: 
 the mean-field behavior  $\rho_{typ}\sim (W_c-W)$ found at
$\lambda=0$ \cite{DobroEPL2003} changes to $(W_c-W)^{1/2}$ at the approach of the MIT,
indicating a radical change in the disorder distribution as soon as $\lambda\neq 0$  \cite{Mahmoudian}.

\paragraph{Self-consistent local potentials.--}
To substantiate this statement,
we introduce the self-consistent field $u=\epsilon+Re \Sigma_{e-ph}(\omega=0,\epsilon)$,
defined as the local energy level renormalized by the interaction with the deformable
lattice. In the static phonon limit  considered first, the real part of the
electron-phonon self-energy reduces  at $T=0$ to the
energy-independent Hartree term
$Re \Sigma(\omega,\epsilon)=\sqrt{\lambda}X_0(\epsilon)$, where $X_0(\epsilon)$ is the static local deformation
on a site, given the local potential $\epsilon$ \cite{MillisPRB1996}.
It is clear from Eq. (\ref{eq:H}) that 
the site disorder acts as a polarization
field coupled to the charge, in full analogy with the external magnetic field in
the Ising model.
Accordingly, an order parameter for the polaron transition
can be  defined as the value $X_0=\lim_{\epsilon\rightarrow 0^+}
X_0(\epsilon)$ much like the remnant magnetization in a ferromagnet, as shown in Fig. 3(a).

\begin{figure} 
\centering
\includegraphics[width= \columnwidth]{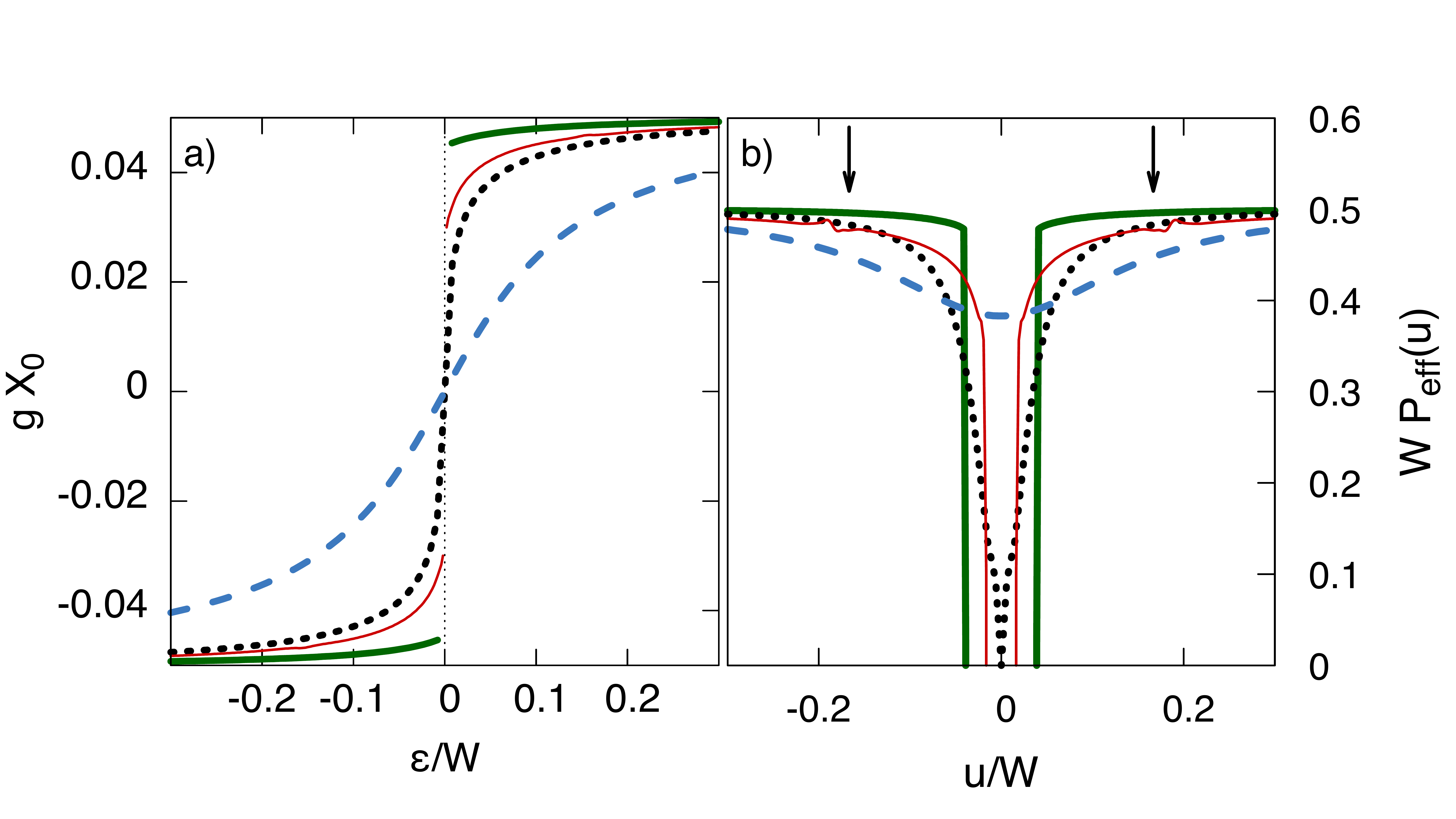}
\caption{\textit{Self-consistent fields and effective disorder.}
(a) Lattice displacement $X_0$ as a function of $\epsilon$ and (b) Effective disorder distribution:
below ($W=1.0$, dashed), at ($W=1.15$, dotted) and beyond the polaronic transition
($W=1.2$, bold), for $\lambda=0.05$ and $T=0$.
Thin lines in (a) and (b) are the NCA results for quantum phonons at $W=1.2$ and $\omega_0=0.2$
(the value of $\omega_0$ is marked by arrows in (b)).
}
\label{fig:X0Peff}
\end{figure}

Inverting for ${\epsilon}(u)$  leads to the effective disorder distribution
$P_{eff}(u)=  P_0({\epsilon}(u))/|1+\frac{\partial \Sigma_{e-ph}}{\partial \epsilon}|$
reported in Fig. \ref{fig:X0Peff}(b), showing that the action of the lattice degrees of freedom
dramatically changes the nature of the disorder. As randomness increases, the presence of
correlated electron-lattice displacements leads to a discontinuity in $X_0(\epsilon)$, signalling
a polaron transition; correspondingly, a
gap opens in $P_{eff}(u)$.
Moreover, the buildup of local deformations correlated with the large fluctuations of the site potentials 
starts already well before the transition,
This causes a dip in the distribution (dashed line in Fig. 3(b)) and a  suppression
of the mobile states available at the Fermi energy (Fig. 2(c)), which has fundamental 
consequences for charge transport as we show next. 

\paragraph{Minimum metallic conductivity.---}

We evaluate the electrical conductivity from the Kubo formula following Refs. \cite{Girvin,AbouChakra},  by 
expressing the current-current correlation function as
$\chi_{JJ}(\omega)=\Lambda(\omega)P_1(\omega)$,
which isolates the dominant non-local contribution $P_1$ responsible for localization.
From Ref. \cite{AlvermanPSSC2004} we have that  $P_1(\omega)=B(\omega)\rho_{typ}(\omega)$
with $B(\omega)$ a weakly $\omega$-dependent function,
so that the conductivity is correctly
proportional to the order parameter of the Anderson transition.
The  prefactor $\Lambda$ is non-critical and
can be calculated within DMFT-CPA as
$\Lambda(\omega)=\chi^{CPA}_{JJ}(\omega)/P^{CPA}_1(\omega)$, leading to 
the following interpolation formula (see Supplemental Material):
\begin{equation}
\sigma=\sigma_0\int d\omega \left ( -\frac{\partial f}{\partial \omega}\right)
\frac{\chi^{CPA}_{JJ}(\omega)}{\rho(\omega)} \rho_{typ}(\omega),
\label{eq:condTMT}
\end{equation}
where $f$ is the Fermi  function and 
$\sigma_0=\pi e^2a^2/\hbar v$ the conductivity unit
($a$ and $v$ are respectively
the lattice parameter and the unit cell volume).
Eq. (2)
can be greatly simplified by taking the $T\to 0$ limit and
introducing the transport scattering time from the semi-classical expression
$ \chi^{CPA}_{JJ}(0)\propto\rho(0) \tau$
\cite{MillisA15}.
The resulting
\begin{equation}
\sigma\propto
\rho_{typ}(0)\ \tau
\label{eq:condTMT0}
\end{equation}
acquires a transparent physical meaning: upon approaching the Anderson insulator, part of the carriers localize
due to quantum interference effects and drop out of the conductivity, which is encoded in $\rho_{typ}$ \cite{Anderson};
the remaining itinerant carriers  are not affected by localization and are therefore
scattered by disorder and  lattice fluctuations in a way that is properly described by the
semi-classical $\tau$.

\begin{figure}
\includegraphics[width=0.9\columnwidth]{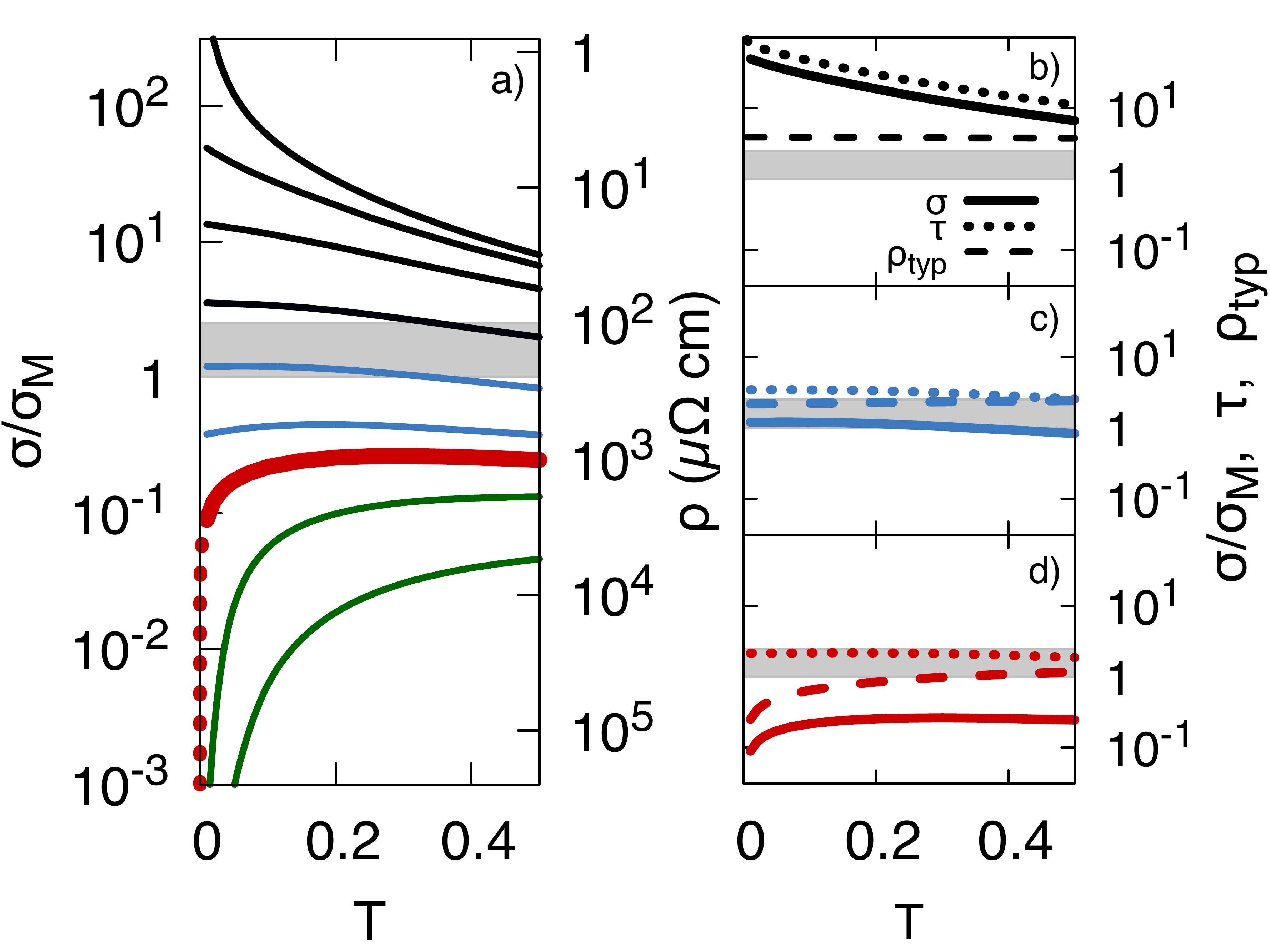}
\caption{\textit{Conductivity and Mott limit.} (a) $\sigma(T)$  
in the weak coupling regime, $\lambda=0.05$, expressed in units of $\sigma_M$ (see text);
from top to bottom, $W=0.0, 0.125, 0.25, 0.5; 0.8, 1.05, 1.16$, $1.3, 1.5$.
For $W=W_c=1.16$ (bold) we also show the power-law extrapolation to $T=0$ (dotted).
The shaded area is the  experimental range  $100 - 250 \mu \Omega cm$ where the
 TCR is seen to vanish in most single-band materials \cite{Tsuei,Hussey}.
(b-d) Scattering time and order parameter for representative parameters in the metallic phase ($W=0.125$), at the Mott-Ioffe-Regel limit
($W=W^*=0.8$) and at the MIT ($W=W_c$). Both quantities are expressed in units of $1/D$.
}
\label{fig:cond}
\end{figure}

The conductivity obtained from Eq. (\ref{eq:condTMT})
is illustrated in Fig. \ref{fig:cond}(a) (note that Eq. (3) would provide essentially indistinguishable
results for $T\lesssim 0.1$). Within the metallic regime at low disorder, the standard Drude-Boltzmann 
picture applies, leading to a conductivity that decreases with 
temperature: this is due to strongly temperature-dependent scattering between (largely) $T$-independent
electronic states. This can be checked directly  in Fig. \ref{fig:cond}(b), which
reports  the behavior of $\tau$ and $\rho_{typ}$ separately.
Upon increasing the disorder strength, the scattering rate progressively increases
($\tau$ decreases) until
it becomes comparable with the bandwidth $D$, cf. Fig. \ref{fig:cond}(c).
At this point,  denoted as $W=W^*$, all quantities including
the conductivity become essentially temperature independent. 
For even stronger disorder, the scattering time cannot be reduced further
as it has already saturated to its minimum value. 

Remarkably, the value of the conductivity
at $W^*$ precisely coincides
with Mott's   minimum metallic conductivity $\sigma_M$, i.e. the Mott-Ioffe-Regel (MIR) limit
\cite{Hussey,GunnarrssonCalandraRMP2003,noteMIR}.
The MIR  limit therefore marks the onset of a regime where transport is not governed by how the electrons are scattered, but rather by the strong $T$-dependence of the
electronic spectrum itself:
a mobility pseudo-gap opens at low $T$  reflecting
disorder-enhanced polaronic processes (cf. Fig. 2(c) and the subsequent discussion), which is progressively filled
upon increasing the temperature as shown in Fig. \ref{fig:cond}(d) (dashed line). Moreover, 
the actual number of mobile 
charge carriers is much smaller than the total number of electrons in the system, as most electronic states 
are now localized.
This results in a "bad insulator"  transport regime (Fig. 4, blue and Fig. 1, shaded) which displays conductivity values
below the  MIR limit and an insulating-like temperature coefficient
$d\sigma/dT>0$ but with a finite d.c. intercept, 
an unexplained behavior that is often observed in strongly disordered metals
\cite{LeeRMP}.
Such intermediate regime ends at the critical point,
where a mobility gap fully opens and the conductivity eventually vanishes at $T=0$
(thick line).

\paragraph{Lattice quantum fluctuations.---}
We now show that the MIT reported in Fig. 1 is a robust phenomenon, i.e.
a genuine transition exists from $\omega_0=0$ all the way to $\omega_0\to \infty$.
The critical line obtained in the static limit  is shown in the inset of Fig. 1. 
In the opposite limit, $\omega_0\to\infty$ ($\lambda$ finite), the 
electron-phonon interaction becomes ineffective for spinless electrons. 
The  transition therefore occurs at the non-interacting value $W_c^{(0)}$
independent of $\lambda$.
We conclude that a MIT must exist for any finite $\omega_0$, 
located between $W_c^{(0)}$ and the critical line  for classical phonons. This is 
indeed what we find numerically (Fig. 1 and Fig. \ref{fig:DOSvsW_DOS}(e), insets).
Increasing values of $\omega_0$ produce a slight stabilization of the metal;
yet, at the coupling strengths attainable with the NCA,
the MIT remains very close to 
the one calculated with classical phonons.
Sizable differences are expected 
instead as one approaches the clean limit $W=0$. 
There, it is known that quantum fluctuations push the MIT to $\lambda\to \infty$,
while the classical limit yields a transition at $\lambda^{(0)}_{c}=0.67$ \cite{MillisPRB1996}. 
A sketch of the behavior inferred from the above considerations 
is shown as a dashed line in Fig. 1.

The key phenomenon that we unveiled,
i.e. the
correlated response of the lattice to the local random potentials, is also preserved at finite $\omega_0$:
as shown in Fig. 3(b), the distribution of the
self-consistent field  is essentially unchanged by phonon quantum fluctuations 
at large $\epsilon$, and the gap in the distribution of $u$
remains finite, although somewhat renormalized as compared to the static phonon case.
In more refined treatments, such gap could be partially filled by exponential tails 
\cite{CaponeCiukPRL2003,pata,Carta}.

\paragraph{Concluding remarks.---}
In this Letter we provided a clear microscopic picture of disorder-driven MITs in deformable lattices, where
disorder-enhanced interaction effects dominate all physical processes. Such interplay is revealed most prominently 
in the emergence of an intermediate regime that separates the conventional metal from the
insulator, where the system still conducts at $T=0$, but it displays resistivity
that {\em decreases} with temperature, i.e.  negative TCR behavior.
Remarkably, the boundary of this anomalous transport regime, found at
intermediate disorder $W=W^*$, is marked precisely by the resistivity reaching
Mott-Ioffe-Regel limit, as argued in very early works by Mott. The actual MIT
point is reached at somewhat stronger disorder $W = W_c > W^*$, and it displays
all signatures of a $T=0$ quantum critical point, with power-law scaling behavior
of all quantities. Our findings, therefore, reconcile Mott's concept of "minimum
metallic conductivity", and the ideas based on the scaling theory of
localization of  Anderson and followers. We showed that both ideas apply, but
they do so in two physically distinct regimes within the phase diagram. Our
results open the road to properly interpret many puzzling experiments in
disordered metals, including the long-standing puzzle of "Mooij Correlations" \cite{LeeRMP,Tsuei},
which remains a challenge for future work.

\paragraph{Acknowledgments ---} V.D. was supported by the NSF grant DMR-1410132. D. Di S. acknowledges
the German Research Foundation (DFG-SFB 1170). S.C. and D. Di S.
acknowledge CINECA ISCRA-C HPC project no. HP10C5W99T and the SuperMUC
system at the Leibniz Supercomputing Centre under the Projekt-ID pr94vu.

\bibliography{Bibliography}

\begin{thebibliography}{40}
\expandafter\ifx\csname natexlab\endcsname\relax\def\natexlab#1{#1}\fi
\expandafter\ifx\csname bibnamefont\endcsname\relax
  \def\bibnamefont#1{#1}\fi
\expandafter\ifx\csname bibfnamefont\endcsname\relax
  \def\bibfnamefont#1{#1}\fi
\expandafter\ifx\csname citenamefont\endcsname\relax
  \def\citenamefont#1{#1}\fi
\expandafter\ifx\csname url\endcsname\relax
  \def\url#1{\texttt{#1}}\fi
\expandafter\ifx\csname urlprefix\endcsname\relax\def\urlprefix{URL }\fi
\providecommand{\bibinfo}[2]{#2}
\providecommand{\eprint}[2][]{\url{#2}}

\bibitem[{\citenamefont{Anderson}(1958)}]{Anderson}
\bibinfo{author}{\bibfnamefont{P.~W.} \bibnamefont{Anderson}},
  \bibinfo{journal}{Phys. Rev.} \textbf{\bibinfo{volume}{109}},
  \bibinfo{pages}{1492} (\bibinfo{year}{1958}).

\bibitem[{\citenamefont{Evers and Mirlin}(2008)}]{MirlinEvers}
\bibinfo{author}{\bibfnamefont{F.}~\bibnamefont{Evers}} \bibnamefont{and}
  \bibinfo{author}{\bibfnamefont{A.~D.} \bibnamefont{Mirlin}},
  \bibinfo{journal}{Rev. Mod. Phys.} \textbf{\bibinfo{volume}{80}},
  \bibinfo{pages}{1355} (\bibinfo{year}{2008}).

\bibitem[{\citenamefont{Lee and Ramakrishnan}(1985)}]{LeeRMP}
\bibinfo{author}{\bibfnamefont{P.~A.} \bibnamefont{Lee}} \bibnamefont{and}
  \bibinfo{author}{\bibfnamefont{T.~V.} \bibnamefont{Ramakrishnan}},
  \bibinfo{journal}{Rev. Mod. Phys.} \textbf{\bibinfo{volume}{57}},
  \bibinfo{pages}{287} (\bibinfo{year}{1985}).

\bibitem[{\citenamefont{Georges et~al.}(1996)\citenamefont{Georges, Kotliar,
  Krauth, and Rozenberg}}]{dmft96}
\bibinfo{author}{\bibfnamefont{A.}~\bibnamefont{Georges}},
  \bibinfo{author}{\bibfnamefont{G.}~\bibnamefont{Kotliar}},
  \bibinfo{author}{\bibfnamefont{W.}~\bibnamefont{Krauth}}, \bibnamefont{and}
  \bibinfo{author}{\bibfnamefont{M.~J.} \bibnamefont{Rozenberg}},
  \bibinfo{journal}{Rev. Mod. Phys.} \textbf{\bibinfo{volume}{68}},
  \bibinfo{pages}{13} (\bibinfo{year}{1996}).

\bibitem[{\citenamefont{Miranda and Dobrosavljevic}(2005)}]{RoP2005review}
\bibinfo{author}{\bibfnamefont{E.}~\bibnamefont{Miranda}} \bibnamefont{and}
  \bibinfo{author}{\bibfnamefont{V.}~\bibnamefont{Dobrosavljevic}},
  \bibinfo{journal}{Reports on Progress in Physics}
  \textbf{\bibinfo{volume}{68}}, \bibinfo{pages}{2337} (\bibinfo{year}{2005}).

\bibitem[{\citenamefont{Pastor and Dobrosavljevi\'c}(1999)}]{pastor-prl99}
\bibinfo{author}{\bibfnamefont{A.~A.} \bibnamefont{Pastor}} \bibnamefont{and}
  \bibinfo{author}{\bibfnamefont{V.}~\bibnamefont{Dobrosavljevi\'c}},
  \bibinfo{journal}{Phys. Rev. Lett.} \textbf{\bibinfo{volume}{83}},
  \bibinfo{pages}{4642} (\bibinfo{year}{1999}).

\bibitem[{\citenamefont{Dobrosavljevi\'c and Kotliar}(1997)}]{motand}
\bibinfo{author}{\bibfnamefont{V.}~\bibnamefont{Dobrosavljevi\'c}}
  \bibnamefont{and} \bibinfo{author}{\bibfnamefont{G.}~\bibnamefont{Kotliar}},
  \bibinfo{journal}{Phys. Rev. Lett.} \textbf{\bibinfo{volume}{78}},
  \bibinfo{pages}{3943} (\bibinfo{year}{1997}).

\bibitem[{\citenamefont{Fisk and Webb}(1976)}]{fisk76prl}
\bibinfo{author}{\bibfnamefont{Z.}~\bibnamefont{Fisk}} \bibnamefont{and}
  \bibinfo{author}{\bibfnamefont{G.~W.} \bibnamefont{Webb}},
  \bibinfo{journal}{Phys. Rev. Lett.} \textbf{\bibinfo{volume}{36}},
  \bibinfo{pages}{1084} (\bibinfo{year}{1976}).

\bibitem[{\citenamefont{Siegrist et~al.}(2011)\citenamefont{Siegrist, Jost,
  Volker, Woda, Merkelbach, Schlockermann, and Wuttig}}]{siwgriest11natphys}
\bibinfo{author}{\bibfnamefont{T.}~\bibnamefont{Siegrist}},
  \bibinfo{author}{\bibfnamefont{P.}~\bibnamefont{Jost}},
  \bibinfo{author}{\bibfnamefont{H.}~\bibnamefont{Volker}},
  \bibinfo{author}{\bibfnamefont{M.}~\bibnamefont{Woda}},
  \bibinfo{author}{\bibfnamefont{P.}~\bibnamefont{Merkelbach}},
  \bibinfo{author}{\bibfnamefont{C.}~\bibnamefont{Schlockermann}},
  \bibnamefont{and} \bibinfo{author}{\bibfnamefont{M.}~\bibnamefont{Wuttig}},
  \bibinfo{journal}{Nat. Mater.} \textbf{\bibinfo{volume}{10}},
  \bibinfo{pages}{202} (\bibinfo{year}{2011}), ISSN \bibinfo{issn}{1476-1122}.

\bibitem[{\citenamefont{Mott}(1990)}]{mott-book90}
\bibinfo{author}{\bibfnamefont{N.~F.} \bibnamefont{Mott}},
  \emph{\bibinfo{title}{Metal-{I}nsulator {T}ransition}}
  (\bibinfo{publisher}{Taylor \& Francis}, \bibinfo{address}{London},
  \bibinfo{year}{1990}).

\bibitem[{\citenamefont{Dobrosavljevi\'c
  et~al.}(2012)\citenamefont{Dobrosavljevi\'c, Trivedi, and
  Valles~Jr.}}]{dobrosavljevic2012conductor}
\bibinfo{author}{\bibfnamefont{V.}~\bibnamefont{Dobrosavljevi\'c}},
  \bibinfo{author}{\bibfnamefont{N.}~\bibnamefont{Trivedi}}, \bibnamefont{and}
  \bibinfo{author}{\bibfnamefont{J.~M.} \bibnamefont{Valles~Jr.}},
  \emph{\bibinfo{title}{Conductor Insulator Quantum Phase Transitions}}
  (\bibinfo{publisher}{Oxford University Press}, \bibinfo{address}{UK},
  \bibinfo{year}{2012}).

\bibitem[{\citenamefont{Anderson}(1972)}]{AndersonNat72}
\bibinfo{author}{\bibfnamefont{P.~W.} \bibnamefont{Anderson}},
  \bibinfo{journal}{Nature} \textbf{\bibinfo{volume}{235}},
  \bibinfo{pages}{163} (\bibinfo{year}{1972}).

\bibitem[{\citenamefont{Shore et~al.}(1973)\citenamefont{Shore, Sander, and
  Kleinman}}]{Shore}
\bibinfo{author}{\bibfnamefont{H.}~\bibnamefont{Shore}},
  \bibinfo{author}{\bibfnamefont{L.}~\bibnamefont{Sander}}, \bibnamefont{and}
  \bibinfo{author}{\bibfnamefont{L.}~\bibnamefont{Kleinman}},
  \bibinfo{journal}{Nature} \textbf{\bibinfo{volume}{245}}, \bibinfo{pages}{44}
  (\bibinfo{year}{1973}).

\bibitem[{\citenamefont{Cohen et~al.}(1983)\citenamefont{Cohen, Economou, and
  Soukoulis}}]{Cohen83}
\bibinfo{author}{\bibfnamefont{M.~H.} \bibnamefont{Cohen}},
  \bibinfo{author}{\bibfnamefont{E.~N.} \bibnamefont{Economou}},
  \bibnamefont{and} \bibinfo{author}{\bibfnamefont{C.~M.}
  \bibnamefont{Soukoulis}}, \bibinfo{journal}{Phys. Rev. Lett.}
  \textbf{\bibinfo{volume}{51}}, \bibinfo{pages}{1202} (\bibinfo{year}{1983}).

\bibitem[{\citenamefont{Shinozuka}(1985)}]{Shinozuka}
\bibinfo{author}{\bibfnamefont{Y.}~\bibnamefont{Shinozuka}},
  \bibinfo{journal}{Journal of Non-Crystalline Solids}
  \textbf{\bibinfo{volume}{77}}, \bibinfo{pages}{21 } (\bibinfo{year}{1985}).

\bibitem[{\citenamefont{Dobrosavljevi\'{c}
  et~al.}(2003)\citenamefont{Dobrosavljevi\'{c}, Pastor, and
  Nikoli\'{c}}}]{DobroEPL2003}
\bibinfo{author}{\bibfnamefont{V.}~\bibnamefont{Dobrosavljevi\'{c}}},
  \bibinfo{author}{\bibfnamefont{A.~A.} \bibnamefont{Pastor}},
  \bibnamefont{and} \bibinfo{author}{\bibfnamefont{B.~K.}
  \bibnamefont{Nikoli\'{c}}}, \bibinfo{journal}{Eur. Phys. Lett.}
  \textbf{\bibinfo{volume}{62}}, \bibinfo{pages}{76} (\bibinfo{year}{2003}).

\bibitem[{\citenamefont{Ekuma et~al.}(2014)\citenamefont{Ekuma, Terletska, Tam,
  Meng, Moreno, and Jarrell}}]{Ekuma14}
\bibinfo{author}{\bibfnamefont{C.~E.} \bibnamefont{Ekuma}},
  \bibinfo{author}{\bibfnamefont{H.}~\bibnamefont{Terletska}},
  \bibinfo{author}{\bibfnamefont{K.-M.} \bibnamefont{Tam}},
  \bibinfo{author}{\bibfnamefont{Z.-Y.} \bibnamefont{Meng}},
  \bibinfo{author}{\bibfnamefont{J.}~\bibnamefont{Moreno}}, \bibnamefont{and}
  \bibinfo{author}{\bibfnamefont{M.}~\bibnamefont{Jarrell}},
  \bibinfo{journal}{Phys. Rev. B} \textbf{\bibinfo{volume}{89}},
  \bibinfo{pages}{081107} (\bibinfo{year}{2014}).

\bibitem[{\citenamefont{Zhang et~al.}(2015)\citenamefont{Zhang, Terletska,
  Moore, Ekuma, Tam, Berlijn, Ku, Moreno, and Jarrell}}]{ZhangJarrell}
\bibinfo{author}{\bibfnamefont{Y.}~\bibnamefont{Zhang}},
  \bibinfo{author}{\bibfnamefont{H.}~\bibnamefont{Terletska}},
  \bibinfo{author}{\bibfnamefont{C.}~\bibnamefont{Moore}},
  \bibinfo{author}{\bibfnamefont{C.}~\bibnamefont{Ekuma}},
  \bibinfo{author}{\bibfnamefont{K.-M.} \bibnamefont{Tam}},
  \bibinfo{author}{\bibfnamefont{T.}~\bibnamefont{Berlijn}},
  \bibinfo{author}{\bibfnamefont{W.}~\bibnamefont{Ku}},
  \bibinfo{author}{\bibfnamefont{J.}~\bibnamefont{Moreno}}, \bibnamefont{and}
  \bibinfo{author}{\bibfnamefont{M.}~\bibnamefont{Jarrell}},
  \bibinfo{journal}{Phys. Rev. B} \textbf{\bibinfo{volume}{92}},
  \bibinfo{pages}{205111} (\bibinfo{year}{2015}).

\bibitem[{\citenamefont{Ciuchi et~al.}(1997)\citenamefont{Ciuchi, de~Pasquale,
  Fratini, and Feinberg}}]{depolarone}
\bibinfo{author}{\bibfnamefont{S.}~\bibnamefont{Ciuchi}},
  \bibinfo{author}{\bibfnamefont{F.}~\bibnamefont{de~Pasquale}},
  \bibinfo{author}{\bibfnamefont{S.}~\bibnamefont{Fratini}}, \bibnamefont{and}
  \bibinfo{author}{\bibfnamefont{D.}~\bibnamefont{Feinberg}},
  \bibinfo{journal}{Phys. Rev. B} \textbf{\bibinfo{volume}{56}},
  \bibinfo{pages}{4494} (\bibinfo{year}{1997}).

\bibitem[{\citenamefont{Fratini and Ciuchi}(2003)}]{rhopolaron03}
\bibinfo{author}{\bibfnamefont{S.}~\bibnamefont{Fratini}} \bibnamefont{and}
  \bibinfo{author}{\bibfnamefont{S.}~\bibnamefont{Ciuchi}},
  \bibinfo{journal}{Phys. Rev. Lett.} \textbf{\bibinfo{volume}{91}},
  \bibinfo{pages}{256403} (\bibinfo{year}{2003}).

\bibitem[{\citenamefont{Millis et~al.}(1999)\citenamefont{Millis, Hu, and
  Das~Sarma}}]{MillisA15}
\bibinfo{author}{\bibfnamefont{A.~J.} \bibnamefont{Millis}},
  \bibinfo{author}{\bibfnamefont{J.}~\bibnamefont{Hu}}, \bibnamefont{and}
  \bibinfo{author}{\bibfnamefont{S.}~\bibnamefont{Das~Sarma}},
  \bibinfo{journal}{Phys. Rev. Lett.} \textbf{\bibinfo{volume}{82}},
  \bibinfo{pages}{2354} (\bibinfo{year}{1999}).

\bibitem[{\citenamefont{Millis et~al.}(1996)\citenamefont{Millis, Mueller, and
  Shraiman}}]{MillisPRB1996}
\bibinfo{author}{\bibfnamefont{A.~J.} \bibnamefont{Millis}},
  \bibinfo{author}{\bibfnamefont{R.}~\bibnamefont{Mueller}}, \bibnamefont{and}
  \bibinfo{author}{\bibfnamefont{B.~I.} \bibnamefont{Shraiman}},
  \bibinfo{journal}{Phys. Rev. B} \textbf{\bibinfo{volume}{54}},
  \bibinfo{pages}{5389} (\bibinfo{year}{1996}).

\bibitem[{\citenamefont{Ciuchi and de~Pasquale}(1999)}]{CiuchiPRB1998}
\bibinfo{author}{\bibfnamefont{S.}~\bibnamefont{Ciuchi}} \bibnamefont{and}
  \bibinfo{author}{\bibfnamefont{F.}~\bibnamefont{de~Pasquale}},
  \bibinfo{journal}{Phys. Rev. B} \textbf{\bibinfo{volume}{59}},
  \bibinfo{pages}{5431} (\bibinfo{year}{1999}).

\bibitem[{\citenamefont{Alben et~al.}(1975)\citenamefont{Alben, Blume,
  Krakauer, and Schwartz}}]{CPAvsExact}
\bibinfo{author}{\bibfnamefont{R.}~\bibnamefont{Alben}},
  \bibinfo{author}{\bibfnamefont{M.}~\bibnamefont{Blume}},
  \bibinfo{author}{\bibfnamefont{H.}~\bibnamefont{Krakauer}}, \bibnamefont{and}
  \bibinfo{author}{\bibfnamefont{L.}~\bibnamefont{Schwartz}},
  \bibinfo{journal}{Phys. Rev. B} \textbf{\bibinfo{volume}{12}},
  \bibinfo{pages}{4090} (\bibinfo{year}{1975}).

\bibitem[{\citenamefont{Alvermann et~al.}(2004)\citenamefont{Alvermann,
  Bronold, and Fehske}}]{AlvermanPSSC2004}
\bibinfo{author}{\bibfnamefont{A.}~\bibnamefont{Alvermann}},
  \bibinfo{author}{\bibfnamefont{F.~X.} \bibnamefont{Bronold}},
  \bibnamefont{and} \bibinfo{author}{\bibfnamefont{H.}~\bibnamefont{Fehske}},
  \bibinfo{journal}{physica status solidi (c)} \textbf{\bibinfo{volume}{1}},
  \bibinfo{pages}{63} (\bibinfo{year}{2004}).

\bibitem[{\citenamefont{Capone and Ciuchi}(2003)}]{CaponeCiukPRL2003}
\bibinfo{author}{\bibfnamefont{M.}~\bibnamefont{Capone}} \bibnamefont{and}
  \bibinfo{author}{\bibfnamefont{S.}~\bibnamefont{Ciuchi}},
  \bibinfo{journal}{Phys. Rev. Lett.} \textbf{\bibinfo{volume}{91}},
  \bibinfo{pages}{186405} (\bibinfo{year}{2003}).

\bibitem[{\citenamefont{Di~Sante and Ciuchi}(2014)}]{NCACPA}
\bibinfo{author}{\bibfnamefont{D.}~\bibnamefont{Di~Sante}} \bibnamefont{and}
  \bibinfo{author}{\bibfnamefont{S.}~\bibnamefont{Ciuchi}},
  \bibinfo{journal}{Phys. Rev. B} \textbf{\bibinfo{volume}{90}},
  \bibinfo{pages}{075111} (\bibinfo{year}{2014}).

\bibitem[{\citenamefont{Nie et~al.}(2015)\citenamefont{Nie, Di~Sante,
  Chatterjee, King, Uchida, Ciuchi, Schlom, and Shen}}]{Sr2TiO4}
\bibinfo{author}{\bibfnamefont{Y.~F.} \bibnamefont{Nie}},
  \bibinfo{author}{\bibfnamefont{D.}~\bibnamefont{Di~Sante}},
  \bibinfo{author}{\bibfnamefont{S.}~\bibnamefont{Chatterjee}},
  \bibinfo{author}{\bibfnamefont{P.~D.~C.} \bibnamefont{King}},
  \bibinfo{author}{\bibfnamefont{M.}~\bibnamefont{Uchida}},
  \bibinfo{author}{\bibfnamefont{S.}~\bibnamefont{Ciuchi}},
  \bibinfo{author}{\bibfnamefont{D.~G.} \bibnamefont{Schlom}},
  \bibnamefont{and} \bibinfo{author}{\bibfnamefont{K.~M.} \bibnamefont{Shen}},
  \bibinfo{journal}{Phys. Rev. Lett.} \textbf{\bibinfo{volume}{115}},
  \bibinfo{pages}{096405} (\bibinfo{year}{2015}).

\bibitem[{\citenamefont{Aguiar et~al.}(2009)\citenamefont{Aguiar,
  Dobrosavljevi\ifmmode~\acute{c}\else \'{c}\fi{}, Abrahams, and
  Kotliar}}]{Aguiar09}
\bibinfo{author}{\bibfnamefont{M.~C.~O.} \bibnamefont{Aguiar}},
  \bibinfo{author}{\bibfnamefont{V.}~\bibnamefont{Dobrosavljevi\ifmmode~\acute{c}\else
  \'{c}\fi{}}}, \bibinfo{author}{\bibfnamefont{E.}~\bibnamefont{Abrahams}},
  \bibnamefont{and} \bibinfo{author}{\bibfnamefont{G.}~\bibnamefont{Kotliar}},
  \bibinfo{journal}{Phys. Rev. Lett.} \textbf{\bibinfo{volume}{102}},
  \bibinfo{pages}{156402} (\bibinfo{year}{2009}).

\bibitem[{\citenamefont{Sangiovanni et~al.}(2012)\citenamefont{Sangiovanni,
  Wissgott, Assaad, Toschi, and Held}}]{SangiovanniAssaad}
\bibinfo{author}{\bibfnamefont{G.}~\bibnamefont{Sangiovanni}},
  \bibinfo{author}{\bibfnamefont{P.}~\bibnamefont{Wissgott}},
  \bibinfo{author}{\bibfnamefont{F.}~\bibnamefont{Assaad}},
  \bibinfo{author}{\bibfnamefont{A.}~\bibnamefont{Toschi}}, \bibnamefont{and}
  \bibinfo{author}{\bibfnamefont{K.}~\bibnamefont{Held}},
  \bibinfo{journal}{Phys. Rev. B} \textbf{\bibinfo{volume}{86}},
  \bibinfo{pages}{035123} (\bibinfo{year}{2012}).

\bibitem[{\citenamefont{Richardella et~al.}(2010)\citenamefont{Richardella,
  Roushan, Mack, Zhou, Huse, Awschalom, and Yazdani}}]{Yazdani}
\bibinfo{author}{\bibfnamefont{A.}~\bibnamefont{Richardella}},
  \bibinfo{author}{\bibfnamefont{P.}~\bibnamefont{Roushan}},
  \bibinfo{author}{\bibfnamefont{S.}~\bibnamefont{Mack}},
  \bibinfo{author}{\bibfnamefont{B.}~\bibnamefont{Zhou}},
  \bibinfo{author}{\bibfnamefont{D.~A.} \bibnamefont{Huse}},
  \bibinfo{author}{\bibfnamefont{D.~D.} \bibnamefont{Awschalom}},
  \bibnamefont{and} \bibinfo{author}{\bibfnamefont{A.}~\bibnamefont{Yazdani}},
  \bibinfo{journal}{Science} \textbf{\bibinfo{volume}{327}},
  \bibinfo{pages}{665} (\bibinfo{year}{2010}).

\bibitem[{\citenamefont{Mahmoudian et~al.}(2015)\citenamefont{Mahmoudian, Tang,
  and Dobrosavljevi\ifmmode~\acute{c}\else \'{c}\fi{}}}]{Mahmoudian}
\bibinfo{author}{\bibfnamefont{S.}~\bibnamefont{Mahmoudian}},
  \bibinfo{author}{\bibfnamefont{S.}~\bibnamefont{Tang}}, \bibnamefont{and}
  \bibinfo{author}{\bibfnamefont{V.}~\bibnamefont{Dobrosavljevi\ifmmode~\acute{c}\else
  \'{c}\fi{}}}, \bibinfo{journal}{Phys. Rev. B} \textbf{\bibinfo{volume}{92}},
  \bibinfo{pages}{144202} (\bibinfo{year}{2015}).

\bibitem[{\citenamefont{Girvin and Jonson}(1980)}]{Girvin}
\bibinfo{author}{\bibfnamefont{S.~M.} \bibnamefont{Girvin}} \bibnamefont{and}
  \bibinfo{author}{\bibfnamefont{M.}~\bibnamefont{Jonson}},
  \bibinfo{journal}{Phys. Rev. B} \textbf{\bibinfo{volume}{22}},
  \bibinfo{pages}{3583} (\bibinfo{year}{1980}).

\bibitem[{\citenamefont{Abou-Chacra et~al.}(1973)\citenamefont{Abou-Chacra,
  Thouless, and Anderson}}]{AbouChakra}
\bibinfo{author}{\bibfnamefont{R.}~\bibnamefont{Abou-Chacra}},
  \bibinfo{author}{\bibfnamefont{D.~J.} \bibnamefont{Thouless}},
  \bibnamefont{and} \bibinfo{author}{\bibfnamefont{P.~W.}
  \bibnamefont{Anderson}}, \bibinfo{journal}{Journal of Physics C: Solid State
  Physics} \textbf{\bibinfo{volume}{6}}, \bibinfo{pages}{1734}
  (\bibinfo{year}{1973}).

\bibitem[{\citenamefont{Tsuei}(1986)}]{Tsuei}
\bibinfo{author}{\bibfnamefont{C.~C.} \bibnamefont{Tsuei}},
  \bibinfo{journal}{Phys. Rev. Lett.} \textbf{\bibinfo{volume}{57}},
  \bibinfo{pages}{1943} (\bibinfo{year}{1986}).

\bibitem[{\citenamefont{Hussey et~al.}(2004)\citenamefont{Hussey, Takenaka, and
  Takagi}}]{Hussey}
\bibinfo{author}{\bibfnamefont{N.}~\bibnamefont{Hussey}},
  \bibinfo{author}{\bibfnamefont{K.}~\bibnamefont{Takenaka}}, \bibnamefont{and}
  \bibinfo{author}{\bibfnamefont{H.}~\bibnamefont{Takagi}},
  \bibinfo{journal}{Philosophical Magazine} \textbf{\bibinfo{volume}{84}},
  \bibinfo{pages}{2847} (\bibinfo{year}{2004}).

\bibitem[{\citenamefont{Gunnarsson et~al.}(2003)\citenamefont{Gunnarsson,
  Calandra, and Han}}]{GunnarrssonCalandraRMP2003}
\bibinfo{author}{\bibfnamefont{O.}~\bibnamefont{Gunnarsson}},
  \bibinfo{author}{\bibfnamefont{M.}~\bibnamefont{Calandra}}, \bibnamefont{and}
  \bibinfo{author}{\bibfnamefont{J.~E.} \bibnamefont{Han}},
  \bibinfo{journal}{Rev. Mod. Phys.} \textbf{\bibinfo{volume}{75}},
  \bibinfo{pages}{1085} (\bibinfo{year}{2003}).

\bibitem[{not()}]{noteMIR}
\bibinfo{note}{We define the Mott limit $\sigma_{M}$ as usual as the value of
  the Drude conductivity when the mean-free-path $\ell$ equals the lattice
  spacing $a$. In a cubic lattice, this corresponds to a scattering rate
  $\hbar/\tau=D/3$. For a concentration $x=1/2$ of spinless electrons as
  considered in this work we obtain in our units $\sigma_{M}=\sigma_0/(2\pi)$.
  The conductivity $\sigma_0$ is fixed by taking a representative value
  $a=3$\AA.}

\bibitem[{\citenamefont{Benedetti and Zeyher}(1998)}]{pata}
\bibinfo{author}{\bibfnamefont{P.}~\bibnamefont{Benedetti}} \bibnamefont{and}
  \bibinfo{author}{\bibfnamefont{R.}~\bibnamefont{Zeyher}},
  \bibinfo{journal}{Phys. Rev. B} \textbf{\bibinfo{volume}{58}},
  \bibinfo{pages}{14320} (\bibinfo{year}{1998}).

\bibitem[{\citenamefont{Capone et~al.}(2006)\citenamefont{Capone, Carta, and
  Ciuchi}}]{Carta}
\bibinfo{author}{\bibfnamefont{M.}~\bibnamefont{Capone}},
  \bibinfo{author}{\bibfnamefont{P.}~\bibnamefont{Carta}}, \bibnamefont{and}
  \bibinfo{author}{\bibfnamefont{S.}~\bibnamefont{Ciuchi}},
  \bibinfo{journal}{Phys. Rev. B} \textbf{\bibinfo{volume}{74}},
  \bibinfo{pages}{045106} (\bibinfo{year}{2006}).

\end{thebibliography}

\end{document}